\documentclass[12pt]{revtex4}

\usepackage{enumerate} 
\usepackage{amsmath} 
\usepackage{amsthm} 
\usepackage{amsfonts} 
\usepackage{amssymb} 
\usepackage{amsxtra}  
\usepackage{amssymb} 
\usepackage{delarray} 
\usepackage{graphicx} 
\usepackage{dcolumn} 
\usepackage{epsfig} 
\usepackage{float} 
\usepackage{hhline} 
\restylefloat{figure} 
\restylefloat{table}

\begin{document}

\title%[The dynamical properties of the crust of neutron stars derived from the realistic NN interactions] %JPG
{Dynamical properties of the crust of neutron stars derived from realistic NN interactions}
\author{P. Grygorov, P. G\"{o}gelein  and H. M\"{u}ther}
\address{Institut f\"{u}r Theoretische Physik, \\
Universit\"{a}t T\"{u}bingen, D-72076 T\"{u}bingen, Germany}
%\ead{grygorov@tphys.physik.uni-tuebingen.de}

\begin{abstract}
The mean free path of neutrino in charged and neutral current reactions is
calculated for inhomogeneous nuclear matter which is expected to appear 
in the crust of neutron stars.  The relevant cross section depends on
Fermi and Gamow-Teller strength distributions, which are derived from the
large-scale shell model calculations within the self-consistent
Skyrme-Hartree-Fock approach and in a relativistic mean-field model. The
inhomogeneous nuclear matter is described in terms of cubic Wigner-Seitz
cells, which allows for a microscopic description of the structures in the
so-called pasta phase of nuclear configurations and provides a smooth
transition to the limit of homogeneous matter. The influence of pasta
phase, its microscopical structure  and geometrical shapes  on neutrino
propagation is considered.

\end{abstract}

%\keywords{Neutron star crust, finite temperature, pairing correlations,
%density dependent relativistic mean field approach, mean free path}

%\pacs{21.60.Jz, 21.65.+f, 26.60.+c, 97.60.Jd}
%\submitto{\JPG}

\maketitle

\section{Introduction}
The transport properties of neutrino play an essential role in the
physics of supernovae core collapse and in the evolution of the
newly born neutron stars. The most important ingredient of
neutrino propagation calculations is the neutrino opacity in a
wide range of densities. Both the charged current (CC) absorption and
neutral current (NC) scattering reactions are important sources of the
neutrino opacity.
In earlier works on neutrino interactions with the homogeneous
nuclear matter the noninteracting baryons were considered
\cite{tubbs75}. Later the strong interaction was
taking into account both in the non-relativistic and relativistic
calculations (see, e.g., \cite{sawyer75, iwamoto82, 
reddy98D, reddy99C, navarro99, margueron03} and ref.
therein). It was shown that the neutrino opacities of interacting matter
may significantly altered from those for the noninteracting case \cite{reddy98D}.  
However, the use of homogeneous matter is a
good approximation, while the existence of the stable quasi-nuclei
in the crust of neutron stars is energetically favorable and must
be taken into account.

At low densities, %($\rho\simeq 0.1\rho_0 \div \rho_0$)
the nuclei in matter are expected to form the Coulomb lattice
embedded in the neutron sea, that minimizes the Coulomb repulsion
between the protons. With increase of density the nuclear pasta
structures occur and the stable nuclear shape may change form from
spherical droplet to rod, slab, tube and bubble shapes
\cite{ravenhall83}. Roughly speaking, the favorable nuclear shape
is determined by a balance between the surface and Coulomb
energies. In the following under "pasta phase" we will assume
quasi-nuclear structures with spherical as well as non-spherical shape, which
are embedded in a neutron sea.

%The appearance of the inhomogeneous phase leads to a
%modification of the equation of state (EOS) of nuclear matter
%\cite{shen98} and has an influence on neutron star cooling
%\cite{yakovlev04, page06}.

Various attempts have been made to describe the ground-state structure of 
pasta phase based on Thomas-Fermi approximation \cite{oyamatsu93}, 
Quantum Molecular Dynamics \cite{Horowitz, caballero06, watanabe03, sonoda07}, 
Hartree-Fock and Relativistic Mean-Field calculations (RMF) within 
the Wigner-Seitz (WS) cell approximation \cite{goegel76, GoegeleinPhD, goegel77}. 
Later the dynamical properties of pasta such as the response 
function and neutrino
mean free path (NMFP) were investigated \cite{Horowitz,sonoda07,Reddy,Burrows}.
It was found that the coherent scattering of neutrinos on inhomogeneous matter 
significantly reduces
the mean free path.   
The collective modes in neutron rich skin of pasta were also calculated 
in a random phase approximation by using spherical 
Skyrme-Hartree-Fock (SHF) method \cite{Khan04}.
It was shown the appearance of the super-giant resonance mode 
at low excitation energy, which may affect the specific heat of the crust of neutron stars. 

The calculation of the NMFP in pasta phase presented in this work
is based on Hartree-Fock calculations in cubic WS cell
\cite{goegel76, GoegeleinPhD, goegel77}, which allows for the
description of non-spherical quasi-nuclear structures such as rods
or slabs and contains the limit of homogeneous matter in a natural
way. The self-consistent calculations are performed for
$\beta$-stable matter in a density range for which the
quasi-nuclear structures discussed above are expected to appear.
For the nuclear Hamiltonian we consider Skyrme forces (SLy$4$) but also
perform calculations within the relativistic mean-field (Hartree)
approximation. The stability of the
pasta phase with increase of the temperature is also discussed.

The NMFP is extracted from the relevant cross sections of neutrino
on different pasta structures.  We pay special attention to the dependence 
of our results on 
the internal structure of the pasta phase and its geometrical shapes. 
The mean free paths obtained from these inhomogeneous structures are compared
with those calculated for homogeneous nuclear matter 
at the same global density, thus one can estimate the influence of the 
inhomogeneous phase on the propagation of neutrinos.

After this introduction the details of Skyrme-Hartree-Fock 
approach with 
pairing will be outlined and a method to calculate the NMFP will
be reviewed in section 2. In section 3 we discuss the density dependent
relativistic mean-field (DDRMF) model. The numerical results are discussed 
in section 4 and the final section 5 contains the main conclusions.

\section{Skyrme-Hartree-Fock calculations}

\subsection{Energy functional}

The Skyrme-Hartree-Fock approach has frequently been described
in the literature \cite{sk1,sk2,bv81,NMB:Ring80}. Therefore we
will outline here only a few basic equations, which will define
the nomenclature. The Skyrme model is defined in terms of an
energy density $\mathcal{H} (\boldsymbol{r})$, which can be split
into various contributions\cite{sk2, Chabanat98}
\begin{equation}
 \mathcal{H} =  \mathcal{H}_K + \mathcal{H}_0 + \mathcal{H}_3
        + \mathcal{H}_{\text{eff}} + \mathcal{H}_{\text{fin}}
        + \mathcal{H}_{\text{so}}
        + \mathcal{H}_{\text{Coul}}, \label{eq:sk1}
\end{equation}
where $ \mathcal{H}_K $ is the kinetic energy term, $
\mathcal{H}_0 $ a zero range term, $ \mathcal{H}_3 $ a density
dependent term, $\mathcal{H}_{\text{eff}} $ an effective mass
term, $ \mathcal{H}_{\text{fin}} $ a finite range term and
$\mathcal{H}_{\text{so}} $ a spin-orbit term. These terms are
given by
\begin{eqnarray}
\mathcal{H}_K
    & = & \frac{ \hbar^2}{2m} \tau,  \notag \\
\mathcal{H}_0
    & = & \textstyle{\frac{1}{4}} t_0
          \big[ (2+x_0) \rho^2 - ( 2x_0 + 1 ) (\rho_p^2 + \rho_n^2 ) \big],  \notag \\
\mathcal{H}_3
    & = & \textstyle{\frac{1}{24}} t_3 \rho^\alpha
          \big[ (2 + x_3 ) \rho^2 - ( 2x_3 + 1 ) ( \rho_p^2 + \rho_n^2 ) \big], \notag \\
\mathcal{H}_{\text{eff}}
    & = & \textstyle{\frac{1}{8}}
        \big[ t_1 (2 + x_1 ) + t_2 (2 + x_2 ) \big] \tau \rho \notag\\
    &   & + \textstyle{\frac{1}{8}} \big[ t_2 ( 2x_2 +1 ) - t_1 ( 2 x_1 + 1 ) \big]
        \big[ \tau_p \rho_p + \tau_n \rho_n \big], \notag \\
\mathcal{H}_{\text{fin}}
    & = & -\textstyle{\frac{1}{32}} \big[3 t_1 ( 2 + x_1 ) - t_2 (2 + x_2 ) \big]
        \rho \Delta \rho \notag \\
    & & + \textstyle{\frac{1}{32}} \big[ 3t_1(2x_1 + 1) + t_2(2x_2 + 1 ) \big]
        \big[\rho_p \Delta \rho_p  +\rho_n \Delta \rho_n  \big], \notag \\
\mathcal{H}_{\text{so}}
    & = & - \textstyle{\frac{1}{2}} W_0
    \big[ \rho \, \boldsymbol{\nabla}\boldsymbol{J}
          +  \rho_p \, \boldsymbol{\nabla}\boldsymbol{J}_p
          +  \rho_n \, \boldsymbol{\nabla}\boldsymbol{J}_n  \big].
          \label{eq:sk2}
\end{eqnarray}
The coefficients $t_i$, $x_i$, $W_0$, and $\alpha$ are the
parameters of a generalized Skyrme force \cite{TD:Bonche85}. 
The imaginary time step was used to solve the Hartree-Fock equations
\cite{Davies80, goegel76, GoegeleinPhD}.
The calculations were performed in a cubic Wigner-Seitz cell with a size of
typically  $20$ fm for charge neutral matter containing protons, neutrons and
electrons in  $\beta$-equilibrium. 
Pairing correlations are included in terms of the BCS approximation 
by assuming a density-dependent zero-range pairing force, 
which has been used in earlier calculations 
\cite{goegel76,goegel77,Montani:2004}.

\begin{figure}
\begin{center}
\mbox{  \includegraphics[width =7cm]{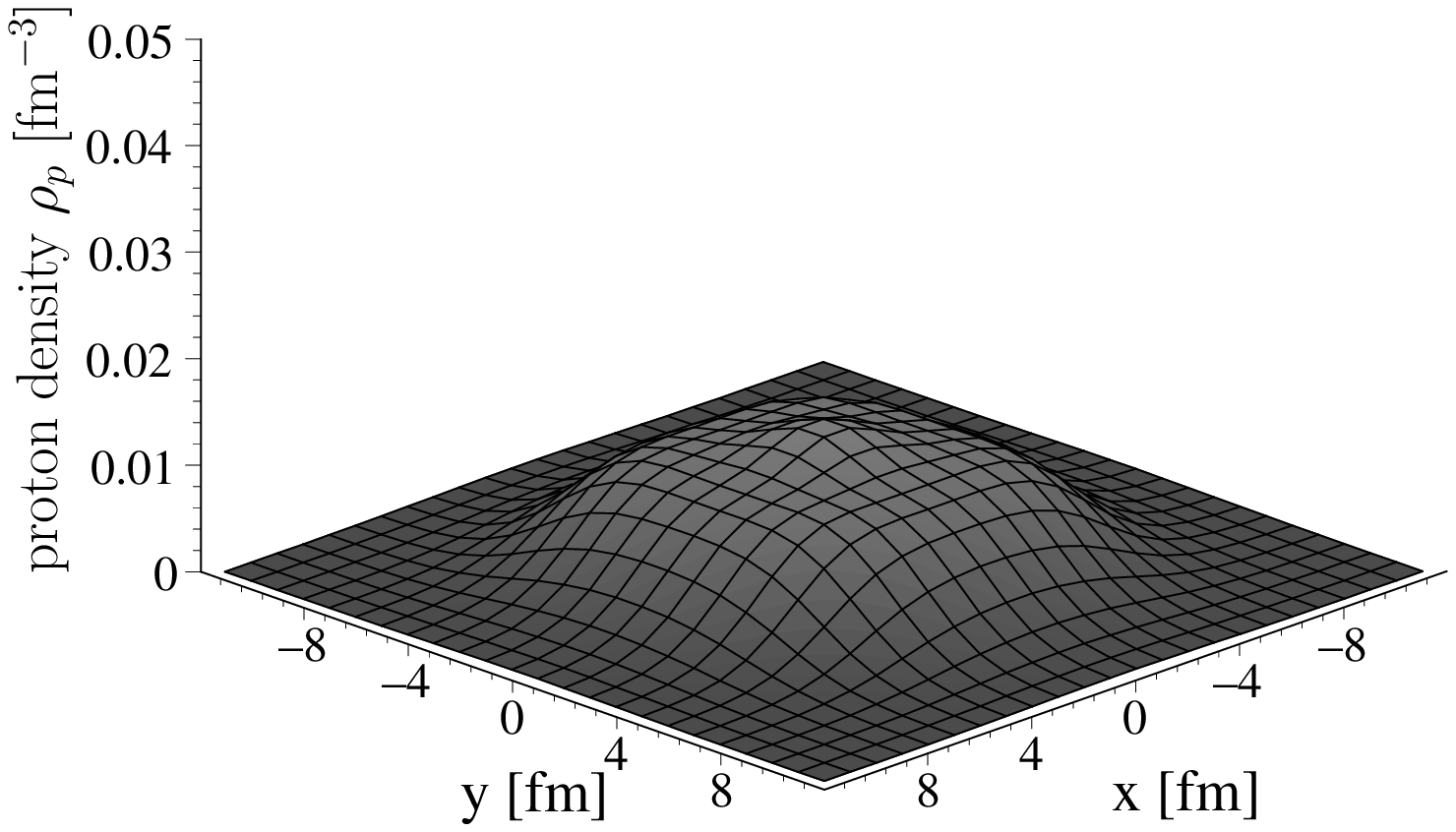}
\includegraphics [width =7cm]{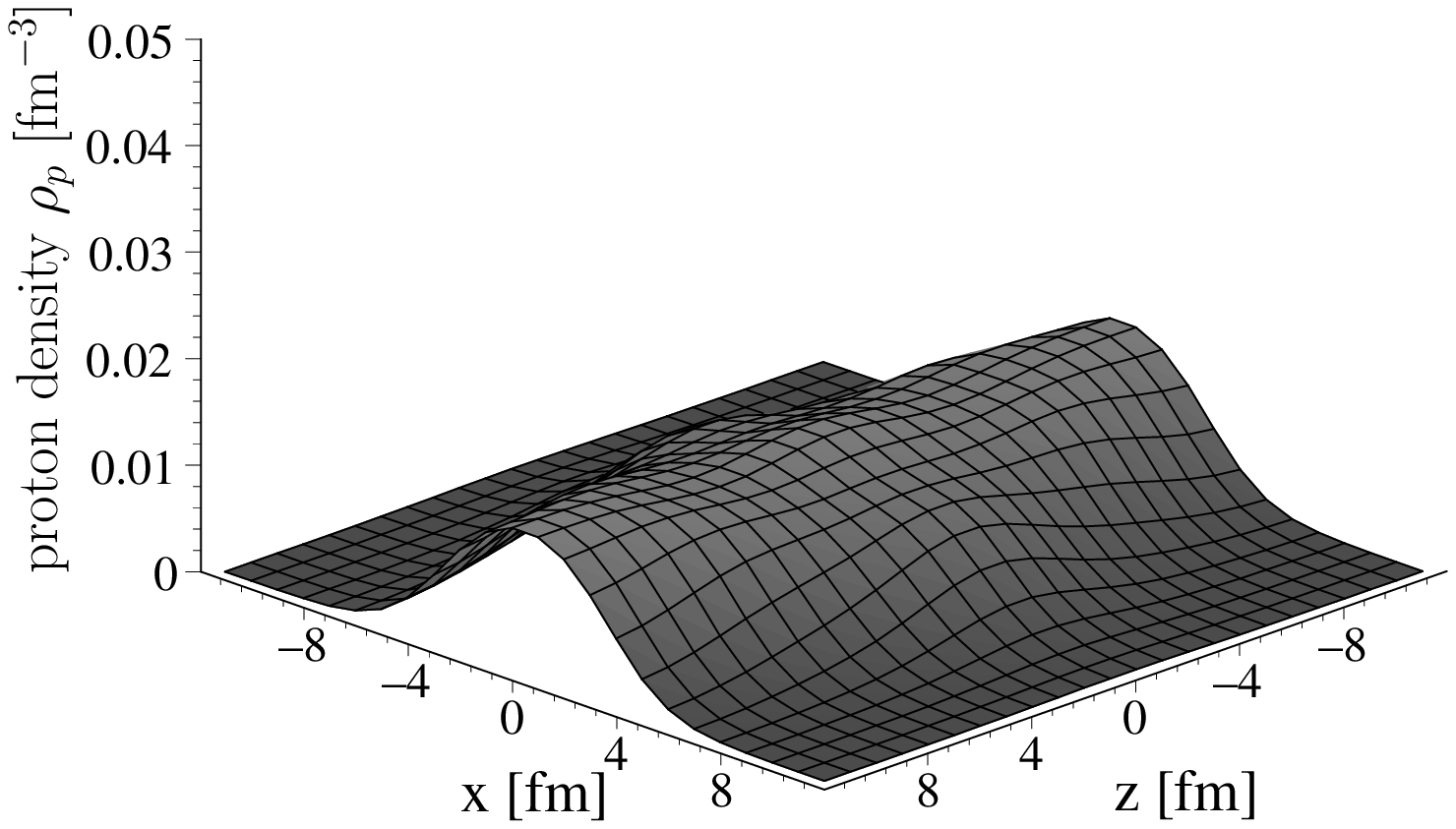}}
\end{center}
\caption{\label{fig:prof1} Profiles for the proton density
distribution forming a rod-structure at a density of 0.0625
fm$^{-3}$.}
\end{figure}

Figure \ref{fig:prof1} displays a typical example for the density profile of the
proton distribution. At the global nucleon density of 0.0625 fm$^{-3}$ we obtain
a quasi-nuclear structure which can be characterized as rods along the z-axis.
At a density of 0.0775 fm$^{-3}$ the variational Hartree-Fock leads to a
quasi-nuclear structure, which can be characterized as a set of parallel slabs.
In figure 2, which shows the proton density distribution at this density, we have 
chosen the orientation of the coordinates such that these slabs are orthogonal
to the z-axis. 
 
The single-particle energies and wave functions for protons and neutrons
resulting from such Hartree-Fock calculations were used
to evaluate the NMFP by using the method outlined in the following subsection.

\begin{figure}
\begin{center}
\mbox{  \includegraphics[width =7cm]{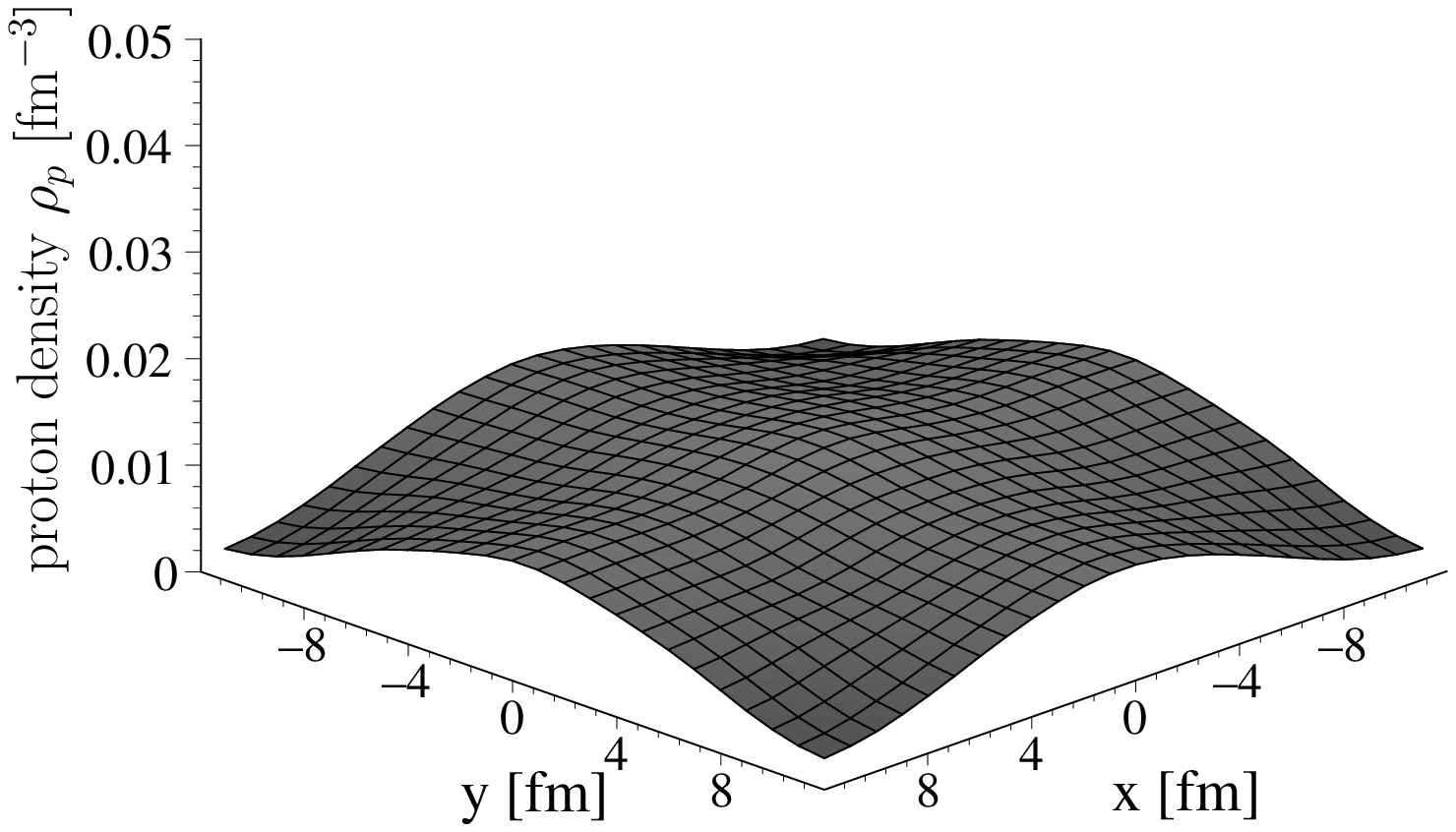}
\includegraphics [width =7cm]{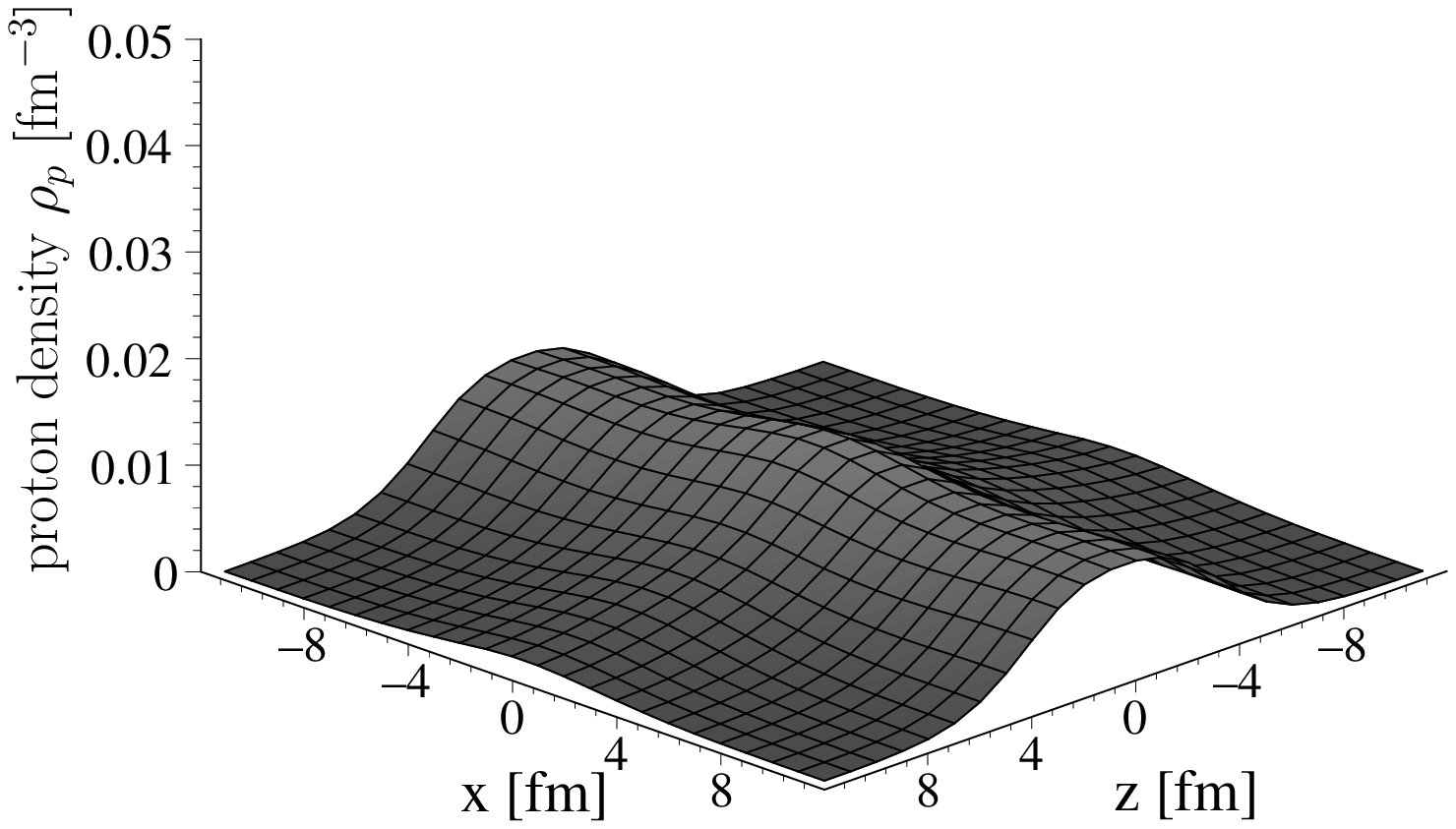}}
\end{center}
\caption{\label{fig:prof2} Profiles for the proton density
distribution forming a slab-structure at a density of 0.0775
fm$^{-3}$.}
\end{figure}

\subsection{Neutrino mean free path in Skyrme-Hartree-Fock model}
 
The matrix element for the neutrino-nucleon reactions
$\nu+n\rightarrow \nu+n$ ( $ \nu+n\rightarrow p+e$ )
%\begin{equation}
%\nu + n \rightarrow e^- + p
%\end{equation}
is given by
\begin{equation}\label{eq:matel}
M = \frac{G_F C}{\sqrt2} J_\mu j^\mu,
\end{equation}
where
\begin{equation}
J_\mu = i\bar{u}_{n(p)} (V\gamma_\mu + A \gamma_\mu\gamma_5)u_n,
\end{equation}
\begin{equation}\label{eq:leptonic}
j^\mu = -i\bar{u}_\nu \gamma^\mu (1-\gamma_5)u_{\nu(e)}
\end{equation}
are hadronic and leptonic currents, respectively. The parameters $A, V$ 
must be replaced by the respective values of coupling constants 
and $C$ stands for the Cabbibo factor in the charged-current reaction \cite{semilepton}.
The total cross section can be written as
\begin{equation}\label{eq:cross_section2}
\sigma = \sum\limits _f p_{\nu(e)} E_{\nu(e)}  \frac12 \int\limits^{1}_{-1}
d(\cos{\vartheta}) \overline{|M|}~^2,
\end{equation}
\begin{equation}\label{eq:matelsquared}
\overline{|M|}~^2=\frac{G_F^2
C^2}{\pi}\left[V^2(1+\cos\vartheta)|M_1|^2 +
A^2(1-\frac{1}{3}\cos\vartheta) |M_2|^2\right],
\end{equation}
\begin{equation}\label{eq:GT}
M_1=\langle\varphi_4| e^{i\vec q \vec r} |\varphi_2\rangle,~~~
M_2=\langle\varphi_4| \vec{\sigma}e^{i\vec q \vec r}
|\varphi_2\rangle,
\end{equation}
where within this non-relativistic approach we neglect the lower components in 
Dirac spinors
$u\simeq\left(^{\varphi}_0 \right)$.
In the charged current reaction the influence of the Coulomb field on the
outgoing electron can be taken into account by multiplying the
cross section by the Fermi function $F(Z_f, E_e)$
\cite{Eisenberg}. For this reaction $M_1$ and $M_2$ stand for the
Fermi and Gamow-Teller matrix elements, respectively.
The integration is performed over the spatial angle $\vartheta$
between the momenta of incoming and the outgoing leptons. The
single-particle wave functions $\varphi(\vec{r})$ and
single-particle energies $\varepsilon_f$ $(\varepsilon_i)$ are
obtained from the solution of the HF equations. Note that these single-particle
energies enter (\ref{eq:cross_section2}) as the energy for the outgoing
lepton is defined as
$$
E_{\nu(e)} = E_{\nu}^{in} +\varepsilon_i - \varepsilon_f\,,
$$
where $E_{\nu}^{in}$ is the energy of the incoming neutrino.

The formalism described so far is appropriate for the neutrino-nucleus 
interaction. With some extensions it may also be used to evaluate the 
interaction of neutrinos with the quasi-nuclear structures in the crust
of neutron stars.
Unlike spherical nuclei and the case of the droplet phase, the cross section 
of neutrino on rods
and slabs, generally speaking, depends on the spatial orientation
of momentum transfer $\vec{q}$ in (\ref{eq:GT}), since the density distributions
of rod and slab phases are non spherical, as it is shown on
figures \ref{fig:prof1},\ref{fig:prof2}. The precise
averaging over all possible mutual orientations of vectors
$\vec{q}$ and $\vec{r}$ requires additional numerical efforts. Thus,
in order to reduce this effort we considered three
particular cases, with the vector $\vec{q}$  along the direction of the
x, y and z-axis. Doing so, we determine the averaged cross section as
\begin{equation}\label{eq:averaged_cross}
\sigma = \frac 13(\sigma_x + \sigma_y + \sigma_z),
\end{equation}
where $\sigma_{x(y,z)}$ represents the cross section calculated for
the momentum transfer along $x (y,z)$-axis.

In contrast to a finite nucleus, the WS cell of the inhomogeneous
nuclear matter contains a large number of unbound neutrons, which
give nonzero contribution to the total cross section. Thus, the
cross section consists of two parts: the cross section due to the interaction
with the nucleons bound in the quasi-nuclear structure
 and the cross section due to the interaction with unbound neutrons.
Therefore, one can consider (\ref{eq:cross_section2}) as a
cross section of neutrinos with all nucleons in a given volume $V_{cell}$ of a 
WS cell. 
The reverse NMFP can then be written as
\begin{equation}\label{eq:NMFP2}
\frac 1\lambda = \frac{\sigma}{V_{cell}}.
\end{equation}
Another important distinction of pasta structures in the crust of neutron stars
from the finite, isolated nuclei consists in the existence of the electron sea 
in the volume of the
WS cell. Therefore the nonzero chemical potential of
electrons must be taken into account in the evaluation of charged current 
reactions by a blocking of final states for electrons with energies
below the respective Fermi energy $\mu_e$.

\section{Relativistic mean-field calculations}

In order to test the sensitivity of the results on the underlying nuclear model
and the
choice of the NN interaction we also investigated the dynamical
properties of inhomogeneous nuclear matter evaluated within a
relativistic mean-field (Hartree) approximation by using a model of
density-dependent meson-nucleon coupling constants. The
parameterization of these constants has been fitted to reproduce the 
properties of the
nucleon self-energy evaluated in Dirac-Brueckner-Hartree-Fock
(DBHF) calculations of asymmetric nuclear matter  but has also been
adjusted to
provide a good description for bulk properties of finite nuclei 
\cite{Klaehn:2006, vanDalen:2007, Schiller:2001, Hofmann:2001}. 
The density-dependent relativistic mean-field (DDRMF) 
approach has also been used to describe the properties of 
inhomogeneous nuclear matter in the crust of neutron stars \cite{goegel77}. 

\subsection{Density dependent relativistic mean-field approach}

% Lagrangian
The  relativistic mean-field (RMF) approach is an effective field
theory of interacting mesons and nucleons. The Lagrangian density consists of three parts: the
free baryon Lagrangian density $\mathcal{L}_B$, the free meson
Lagrangian density $\mathcal{L}_M$ and the interaction Lagrangian
density $\mathcal{L}_{\text{int}}$:
\begin{equation}\label{Lag_dens}
    \mathcal{L} = \mathcal{L}_B + \mathcal{L}_M + \mathcal{L}_{\text{int}},
\end{equation}
which take the explicit form
\begin{equation}
  \begin{split}
  \mathcal{L}_B =\,&  \bar{\Psi} ( \, i \gamma _\mu \partial^\mu - M ) \Psi,  \\
  \mathcal{L}_M =\,&  {\textstyle \frac{1}{2}} \sum_{\iota= \sigma, \delta}
            \Big( \partial_\mu \Phi_\iota \partial^\mu \Phi_\iota - m_\iota^2 \Phi_\iota^2 \Big)   \\
         &  - {\textstyle \frac{1}{2}} \sum_{\kappa = \omega, \rho, \gamma }
            \Big( \textstyle{ \frac{1}{2}} F_{(\kappa) \mu \nu}\, F_{(\kappa)}^{\mu \nu}
                - m_\kappa^2 A_{(\kappa)\mu} A_{(\kappa)}^{\mu} \Big),      \\
  \mathcal{L}_{\text{int}} =\,& - g_\sigma\bar{\Psi}  \Phi_\sigma \Psi
                - g_\delta \bar{\Psi}  \boldsymbol{\tau} \boldsymbol{\Phi}_\delta \Psi \\
        & - g_\omega \bar{\Psi}  \gamma_\mu A_{(\omega)}^{ \mu } \Psi
        - g_\rho \bar{\Psi}  \boldsymbol{\tau } \gamma_\mu  \boldsymbol{A}_{(\rho)}^{\mu } \Psi \\
                & - e \bar{\Psi}\gamma_\mu {\textstyle \frac{1}{2}}(1+ \tau_3 ) A_{(\gamma)}^{\mu} \Psi ,
  \end{split}
\end{equation}
with the field strength tensor
$F_{(\kappa)\mu \nu} = \partial_{\mu} A_{(\kappa)\nu} - \partial_\nu A_{(\kappa)\mu}$
for the vector mesons.
In the above Lagrangian density the nucleon field consisting of Dirac-spinors in
isospin space is denoted by $\Psi$ and the nucleon rest mass by $M = 938.9$ MeV.
The scalar meson fields are $ \Phi_\sigma$ and $ \boldsymbol{\Phi}_\delta $,
the vector meson fields $ A_{(\omega)} $ and $ \boldsymbol{A}_{(\rho)} $.
%Bold symbols denote vectors in the isospin space acting between the two species of nucleons.
The mesons have rest masses $m_\kappa$ for each meson $\kappa$ and
couple to the nucleons with the strength of the coupling constants
$g_\kappa$, which depend on a density of the nucleon field $\Psi$.
A very convenient parameterization for this density dependence
has been given in \cite{vanDalen:2007}.

The numerical procedure to solve the Dirac equation in the cubic
WS cell is the same as in \cite{goegel76, GoegeleinPhD, goegel77}.
Pairing correlations are included in terms of the BCS
approximation assuming a density dependent zero-range pairing
force, which has already been discussed.
The resulting single-particle energies and spinors were used in the
calculation of NMFP as described in the next Subsection.

\subsection{Neutrino mean free path in relativistic mean-field model }

First, let us consider the charged current reaction.  Here we will
exploit the most general form for the nucleonic current, which is
allowed due to the Lorentz, parity and isospin invariances
\cite{semilepton}
\begin{equation}\label{eq:nucloncurrent}
J_\mu^{CC}=i\bar{\psi}_p[F_1^v(q^2)\gamma_\mu+F_2^v(q^2)\sigma_{\mu\nu}q_\nu+F_A(q^2)\gamma_5\gamma_\mu-iF_p(q^2)\gamma_5q_\mu]\psi_n,
\end{equation}
where
$F_1^v$ and $F_2^v$ are isovector electromagnetic formfactors,
$F_A$ is the axial-vector formfactor,  $F_p$ is the induced
pseudoscalar formfactor. %, and $m_{\pi}$ stands for the pion mass.
Following the common practice we ignore the contribution of the
second-class currents. The leptonic current has the same structure
as in (\ref{eq:leptonic}). Analogously to (\ref{eq:matel}),
(\ref{eq:matelsquared}) the averaged squared matrix element for
the charged current reaction can be written in the form
\begin{eqnarray}\label{eq:matelsquare}
\overline{|M|}~^2 & = & \frac{G_F^2 C^2}{2}[~
|\mathcal{M}_1|^2 (1-\frac{p_l}{3E_l}\cos\vartheta)+
|\mathcal{M}_2|^2(1+\frac{p_l}{E_l}\cos\vartheta)
\\
&&
+|\mathcal{M}_3|^2(p_l^2+E_\nu^2-2p_lE_\nu\cos\vartheta-\frac{p_l^3}{3E_l}
\cos\vartheta
-\frac{p_l}{3E_l}E^2_\nu\cos\vartheta
+ \frac{2p_l^2}{3E_l}E_\nu)\nonumber\\
&&
+|\mathcal{M}_4|^2\frac{p_l}{E_l}((p_l^2+E_\nu^2)\cos\vartheta -
2E_lE_\nu\cos\vartheta + p_lE_l+\frac{E_l}{p_l}E_\nu^2 - 2p_l
E_\nu)],\nonumber
\end{eqnarray}
where
\begin{eqnarray}
\mathcal{M}_1 & = &F_1\bar{\psi}_p \vec{\gamma} \psi_n + F_A\bar
\psi_p\gamma_5\vec\gamma \psi_n,\nonumber \\
\mathcal{M}_2 &= &F_1\bar \psi_p \gamma_0 \psi_n + F_A \bar \psi_p
\gamma_5\gamma_0 \psi_n - i F_p \bar \psi_p\gamma_5 q_0 \psi_n,
\nonumber \\
\mathcal{M}_3 &= &F_2\bar \psi_p \vec{\Sigma} \psi_n,
\nonumber \\
\mathcal{M}_4 &= &F_p\bar \psi_p \gamma_5 \psi_n,\nonumber
\end{eqnarray}
and
$$
\vec{\Sigma}=\left(%
\begin{array}{cc}
  \vec{\sigma} & 0            \\
  0            & \vec{\sigma} \\
\end{array}%
\right).
$$

Dirac spinors $\psi$ and the respective single-particle energies
are obtained from the solution of the Dirac equation \cite{goegel77}.

The hadronic part of the neutral current involves additionally
isoscalar electromagnetic form factors $F_1^s$ and $F_2^s$, so
that
\begin{eqnarray}\label{eq:nucleoncurrent2}
J_\mu^{NC} &=& \frac i 2 \bar{\psi}_n[F_A(q^2)\gamma_5\gamma_\mu
    -iF_p(q^2)\gamma_5q_\mu\\
    && +(1-2\sin^2\theta_W)(F_1^v(q^2)\gamma_\mu+F_2^v(q^2)\sigma_{\mu\nu}q_\nu)
    \nonumber\\
    && - 2sin^2\theta_W(F_1^s(q^2) \gamma_\mu + F_2^s(q^2)\sigma_{\mu\nu}q_\nu )]
    \psi_n,\nonumber
\end{eqnarray}
where  $\theta_W$ is the Weinberg angle. The respective matrix
element for this reaction looks like
\begin{eqnarray}\label{eq:matelsquare2}
\overline{|M|}~^2= & \frac{G_F^2}{8}[~ |\mathcal{N}_1|^2
(1-\frac13\cos\vartheta)+ |\mathcal{N}_2|^2(1+\cos\vartheta)\\
 & +|\mathcal{N}_3|^2(E_\nu'^2+E_\nu^2-2E_\nu E_\nu'\cos\vartheta -
\frac 13 E_\nu'^2\cos\vartheta - \frac 13 E_\nu^2\cos\vartheta + \frac
23 E_\nu E_\nu'^2)\nonumber \\
 & +|\mathcal{N}_4|^2(E_\nu - E_\nu')^2(1+\cos\vartheta)],\nonumber
\end{eqnarray}
where
\begin{eqnarray}
\mathcal{N}_1 & = &((1-2\sin^2\theta_W)F_1^v - 2\sin^2\theta_W F_1^s)
\bar{\psi}_n\vec{\gamma}\psi_n +
F_A\bar{\psi}_n\gamma_5\vec{\gamma},\nonumber \\
\mathcal{N}_2 & = & -F_A\bar{\psi}_n\gamma_5\gamma_0\psi_n -
((1-2\sin^2\theta_W)F_1^v-2\sin^2\theta_W F_1^s) \bar{\psi}_n
\psi_n + iF_P\bar{\psi}_n\gamma_5\psi_n,
\nonumber \\
\mathcal{N}_3 & = &((1-2\sin^2\theta_W)F_2^v-2\sin^2\theta_W F_2^s)
\bar{\psi}_n \vec{\Sigma} \psi_n,\nonumber \\
\mathcal{N}_4 & = & F_p\bar \psi_n \gamma_5 \psi_n.\nonumber
\end{eqnarray}

Substituting (\ref{eq:matelsquare}) and (\ref{eq:matelsquare2}) in
(\ref{eq:cross_section2}) one obtains the mean free path of
neutrinos in relativistic mean-field model for charged and
neutral current reactions, respectively.

\section{Results and discussions}

\begin{figure}
\begin{center}
  \mbox{ \includegraphics[width = 11cm]{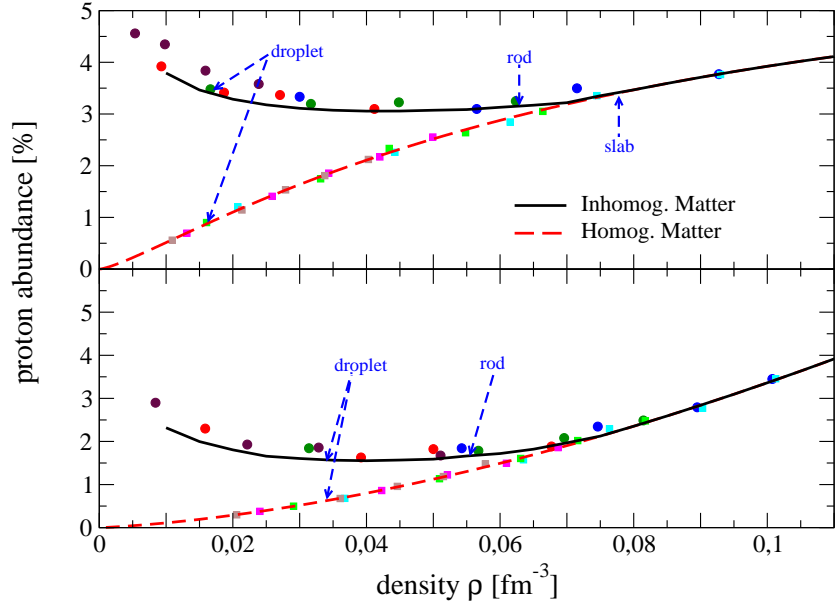} }
\end{center}
\caption{\label{fig:Yp}  Proton abundance in the case
of uniform matter (dashed line) and pasta phase (solid line). 
The symbols refer to specific calculations, whereas the lines have been
added to guide the eye. 
The results for Skyrme-Hartree-Fock calculations are shown in the upper panel
and the relativistic mean-field results in the lower one. The dashed arrows
indicate typical densities leading to pasta structures of droplet, rod and slab
shape.}
\end{figure}

Before we start the discussion of the transport properties of neutrinos in the
crust of neutron stars  let us review some details of self-consistent
Hartree-Fock and relativistic mean-field calculations.  For all the results
presented in this manuscript a temperature $T=1$ MeV was  considered. This
temperature is high enough to take into account some effects of finite
temperatures and low enough to maintain the pairing correlations and stable
quasi-nuclear structures. If the temperature rises,  the pasta phase structures
become smoother and at some critical temperature they disappear. For the
Skyrme-Hartree-Fock approach this critical temperature is around $5$ MeV and
$10$ MeV for slab and rod structures, respectively, while the droplet structure
disappears at a temperature higher than $15$ MeV. Thus, the spherically
symmetric droplet phase will play the main role in different simulations
containing the temperature evolution. Employing the relativistic approach the
pasta phase structures turn out to be less stable and even the droplet
structures disappear at a temperature of $T=10$ MeV. 

For temperatures below 1 MeV and global densities below 0.08 fm$^{-3}$
the variational calculations yield structures with
inhomogeneous density distributions (see figures \ref{fig:prof1} and
\ref{fig:prof2}). Comparing the spectra of single-particle energies obtained for
the homogeneous and inhomogeneous solutions one observes that the single-particle
energies for the localized states are more deeply bound than the corresponding 
single-particle states for the homogeneous approach. In the
$\beta$-equilibrium all proton states are localized and therefore tend to have
more attractive single-particle energies in the inhomogeneous as compared to the
homogeneous density calculation. The variational calculations allowing for pasta
structure yield larger proton fractions than obtained for the
$\beta$-equilibrium of homogeneous matter at the same global density.

This can be seen from inspecting figure \ref{fig:Yp}.  The upper panel of this
figure contains results of the proton abundances for baryonic matter in 
$\beta$-equilibrium resulted from non-relativistic Skyrme-Hartree-Fock
calculations. The proton abundance of homogeneous matter is a monotonically
increasing function of total density and it reaches the value of $4 \%$ at the
density $0.1 fm^{-3}$. Allowing for inhomogeneous matter distribution one
obtains a
significant increase  of the proton fraction at densities below $0.03 fm^{-3}$, 
while in  the density region  from $0.03$ to $0.08 fm^{-3}$  its value is
almost constant around $3.2 \%$. The lower panel of figure \ref{fig:Yp} displays
the corresponding results derived from the relativistic mean field approach.
This relativistic approach seems to provide a smaller symmetry energy at these
low densities, which leads to smaller proton abundances in the inhomogeneous as
well as the homogeneous solution. 

\begin{figure}
\begin{center}
\mbox{ \includegraphics[width = 12cm]{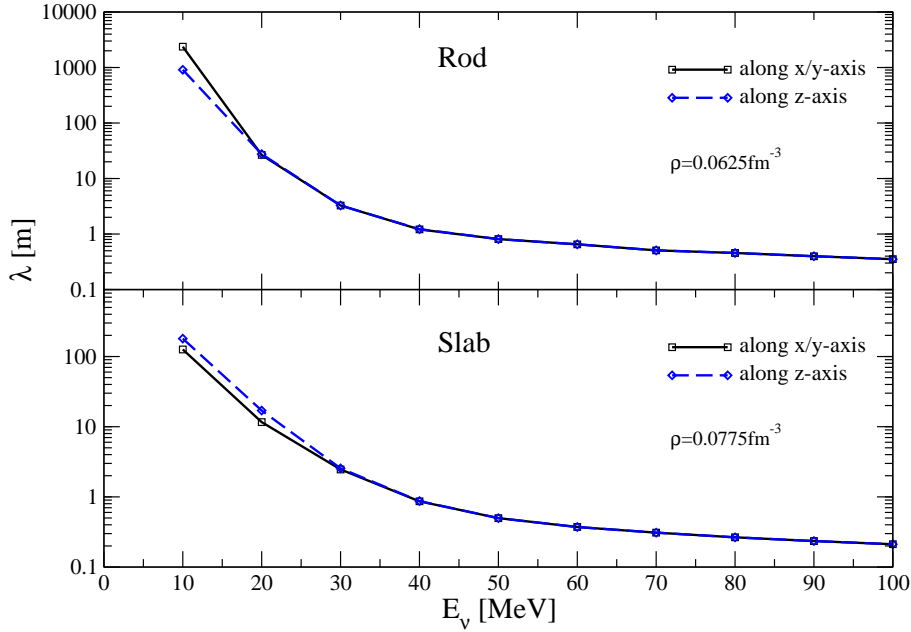} }
\end{center}
\caption{ \label{fig:orientation} The neutrino mean free path (NMFP) calculated
for the charged current reaction in case of rod and slab configurations
demonstrate the dependence of the result on spatial orientation of the momentum
transfer $\vec{q}$.  For these calculations we have employed results of the
Skyrme HF approach and ignore  the blocking of final electron states.}
\end{figure}

In figure \ref{fig:orientation} we want to demonstrate the dependence of the 
neutrino cross section for the charge current reaction on the spatial 
orientation of the momentum transfer $\vec q$. This is displayed in terms of the
corresponding neutrino mean free path, which has been calculated
according to (\ref{eq:NMFP2}) from $\sigma_x$ (solid line) and $\sigma_z$
(dashed line), respectively. Note that due to our choice of the coordinate
system the results for $\sigma_y$ are identical to those for $\sigma_x$ for the
rod as well as the slab structures.

For the density $\rho$ of 0.0625 fm$^{-3}$, which leads to a rod structure, we
obtain results for the NMFP ranging 20 km for neutrinos with an energy of 10 MeV
down to 30 cm for neutrinos with an energy of 100 MeV. For low-energy neutrinos 
the NMFP for reactions with a momentum transfer parallel to the $x$-axis is
larger by a factor of 2 as compared to a momentum transfer parallel to the
$z$-axis, difference which disappears for neutrinos with larger energies. This
factor of 2 is non-negligible but small on the scale of variations for the NMFP
as a function of the neutrino energy. Therefore the simple averaging procedure 
of (\ref{eq:averaged_cross}) seems to be adequate.

Similar results are obtained for the slab configuration as can be seen from the
lower panel of figure \ref{fig:orientation}. Note that the results for the NMFP are
considerably smaller at low neutrino energies (by a factor of 10) and even at
neutrino energies as large as 100 MeV smaller by a factor 2, although the ratio
of the inverse densities is only about 1.2.

\begin{figure}
\begin{center}
\mbox{ \includegraphics[width = 12cm]{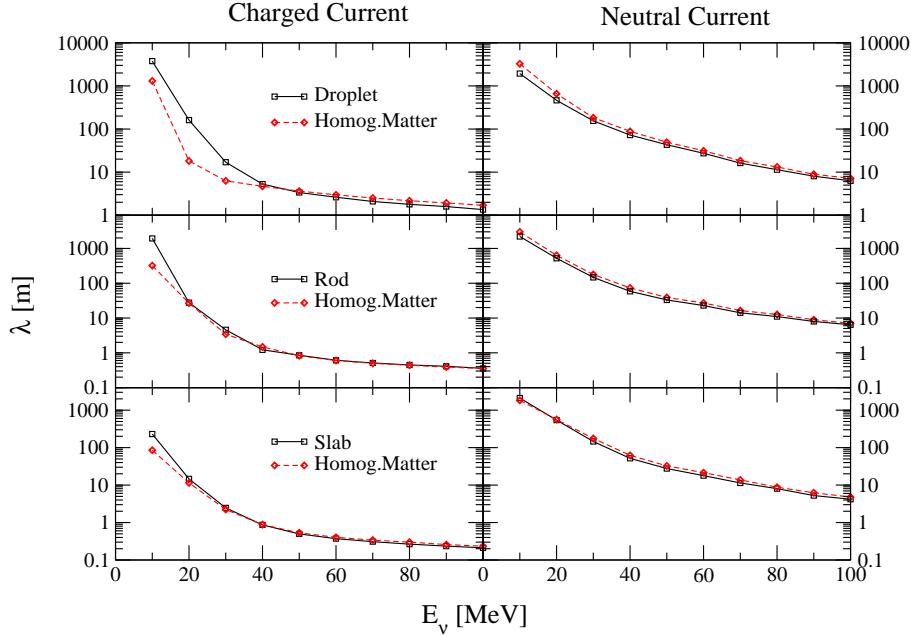} }
\end{center}
\caption{ \label{fig:Sk_3D}  Skyrme-Hartree-Fock
calculations of NMFP for pasta phases (solid curves) and the
respective results for homogeneous matter at the same global
density (dashed curves). The results for the charged current
reaction are shown in the left column, while the neutral current
NMFP in the right one.}
\label{fig:SK_3D_both}
\end{figure}

The NMFP calculated in CC and NC reactions for homogeneous and inhomogeneous
matter distributions  are shown in figure \ref{fig:SK_3D_both}. First, let us
compare the NMFP of homogeneous matter for both types of reactions.  The main
influence on NMFP's results from the available phase space for each reaction. In
fact, the proton fraction of homogeneous matter does not exceed $1\%$ for the
densities considered here. Thus we have to consider a much larger blocking
effect for the neutrons in the final states NC reactions than for the protons in
the CC reactions. Therefore the cross section of CC  absorption is larger than
in NC scattering,  and consequently, the mean free path is shorter, as it is
shown by the red dashed lines in figure \ref{fig:SK_3D_both}. Due to the small
proton abundances in homogeneous matter the Pauli blocking factor of final
electron states affects the result for the CC reaction only at very  small
neutrino energies   $E_\nu < 10$ MeV.

\begin{figure}
\begin{center}
  \mbox{ \includegraphics[width = 11cm]{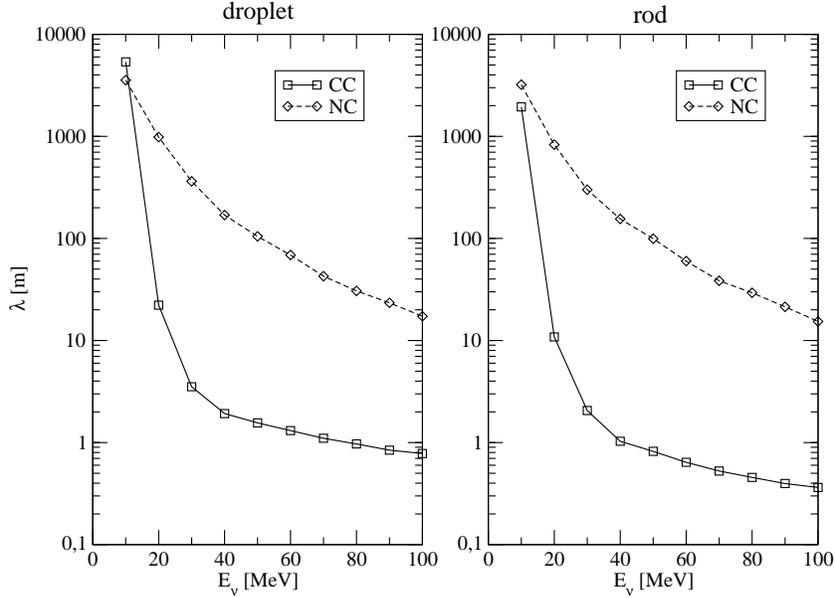} }
\end{center}
\caption{\label{fig:DDRMF}  Results of NMFP due to NC (dashed lines) and CC
(solid lines) reactions. The description of the inhomogeneous baryonic matter
distributions results from the density dependent
relativistic mean-field calculations. As examples we present results for the
droplet phase displayed in the left panel at a density of 0.034 fm$^{-3}$ and 
for the rod phase (right panel) at 0.055 fm$^{-3}$.}
\end{figure}

Figure \ref{fig:SK_3D_both} also presents results for  the NMFP of inhomogeneous 
matter for both types of currents. First of all, we should emphasize the larger
influence of electron blocking factor on CC  current reaction in the droplet
phase. This is due to the larger proton abundances in the $\beta$-equilibrium of
the inhomogeneous matter. At a neutrino energy around $E_\nu\simeq 10$MeV the
mean free path of CC processes  is longer in comparison with  NC scattering,
because in this region  the Pauli blocking of electrons in CC reaction dominates
over the differences in phase spaces of the baryonic states. If the energy of
incoming neutrino $E_\nu$ rises  the Pauli blocking drops exponentially and the 
ratio of the cross sections for CC and NC reactions is determined by the
available phase space for the baryonic states as discussed above for the
homogeneous matter calculation. This means that the NMFP of absorption due to CC
becomes shorter  than the respective result in NC scattering.  At higher
densities, where rods and slabs appear, the influence of Pauli blocking of
electrons is partially  compensated by the effects of the baryonic phase space. 
Therefore the NMFP of CC reaction remains shorter in comparison with NC reaction
for all neutrinos with $10<E_\nu<100$. 

The same features are also observed in
the comparison of NMFP due to the different currents for the models of
inhomogeneous baryonic matter, which are based on the relativistic mean field
calculations displayed in figure \ref{fig:DDRMF}.

The cross section for neutrino scattering in homogeneous matter increases with the
baryonic density in a non-linear way (see discussion above). Therefore one may
expect that the  mean free path in the inhomogeneous matter is shorter than the
corresponding one for homogeneous of the same global density,  since the
scattering on the quasi-nuclear structures shall enhance the respective cross
section. Nevertheless, the NMFP obtained for the charged current reaction, which
is shown in left column of figure \ref{fig:Sk_3D} demonstrates the opposite
behavior, specially at low densities, where the droplet phase occurs. The NMFP
obtained from absorption  in inhomogeneous matter is longer than the respective
result derived from homogeneous matter calculations.

In order to explain this effect one should consider difference in proton
fractions of  homogeneous and inhomogeneous matter discussed in the beginning of
this Section. At a typical density  $0.0165 fm^{-3}$, where the droplet phase
occurs the proton abundance  in inhomogeneous matter is significantly larger
than the respective value obtained in the homogeneous matter. This difference in
the proton fractions has two effects: first, the homogeneous matter contains
less protons in comparison with the inhomogeneous one. Consequently, the number
of unoccupied final proton states is larger and more transitions, which
contribute to the total cross section, are possible. Secondly, the chemical
potential of electrons compensating the charge of the protons in matter is 
lower in case of homogeneous matter and the respective Pauli blocking factor for
the produced electrons is lower than those obtained for the inhomogeneous
matter. This effect again modifies the cross section considerably at low
$E_\nu$. With increase of the energy of incoming neutrinos the Pauli blocking of
electrons rapidly drops and  more transitions become possible, so that the
differences between homogeneous and inhomogeneous matter distributions are
getting less significant and the respective NMFP's become closer one to another.
At higher densities of matter, where the rod and slab phases occur,  the
difference in proton abundances are less important, therefore the resulted mean
free paths are very similar and the effect of inhomogeneous structure becomes
negligible.

At the end we should notice that at neutrino energies less than $10$ MeV
(thermalized neutrinos) the NMFP's of homogeneous and inhomogeneous matter
distributions  calculated in CC reaction significantly exceed the typical
neutron star radius. Therefore one can conclude that the charged current
reaction is  kinematically suppressed \cite{Lattimer91}. 
  
The results of neutral current reaction are shown on the right panel of
figure \ref{fig:Sk_3D}. It is obvious that the appearance of pasta phase in this
case has no important influence on neutrino propagation, since this type of
reaction does not depend on Pauli blocking of neutrino in final state (no
trapped neutrinos). The only small difference in NMFP's of homogeneous matter
and droplet phase may be explained by different values of matrix elements in
(\ref{eq:GT}), since the s.p. wave functions of bound neutrons in droplet
significantly differ from wave functions of homogeneous matter. However even
this small effect becomes negligible if the global density increases and the
density profiles become smoother and  transition to the homogeneous phase
approaches. 

A comparison of NMFP's of charged and neutral currents in case of pasta phase 
based on a relativistic mean-field
 model in a WS cell is displayed in
figure\ref{fig:DDRMF}. It is worth mentioning that within the relativistic model
we could not find any formation of slab structures. Therefore only results for
droplet and rod structures are shown.  Also, the global density, at which the
droplet phase occurs in the relativistic mean-field model is  two times larger
than the respective density in the nonrelativistic model. The difference between
proton fractions of homogeneous matter and pasta phase is not so significant. In
fact, the values of proton abundance around $\rho=0.02$ fm$^{-3}$, displayed in
the lower panel of figure \ref{fig:Yp}, are about $40 \%$ smaller than the
corresponding values obtained in the Skyrme model (the upper panel).  Therefore
we omit the comparison between NMFP's of homogeneous and inhomogeneous matter
however we compare the mean free paths of pasta phase for both types of
reactions. One can see that at $E_\nu < 20$MeV the behavior of CC curves is 
determined by the Pauli blocking, while at higher energies the result becomes
sensitive to the structure of phase space available  for the reactions. Both
charged and neutral current mean free paths  decrease if the global density of
matter rises.

Summarizing we conclude that the NMFP is determined by two different factors.
The first of them - the Pauli blocking effect of final electrons in CC reaction
play the most  important role at low neutrino energy and drops exponentially if
the energy increases. The second factor is the difference in baryonic phase
spaces of different reactions. The phase space of CC absorption is larger than
in NC scattering, because the Fermi energy of final (proton) states is
considerably lower than the neutron Fermi energy. 
 
\section{Summary and Conclusion}
The aim of this study was to examine the role of the inhomogeneous baryonic
density structures in the crust of neutron stars on the 
propagation of neutrinos.
Our calculations of neutrino mean free paths (NMFP) are based on microscopic 
descriptions of the so-called ``pasta structures'' derived
from 3D Hartree-Fock calculations with  the SLy4 parameterization of the Skyrme 
potential as well as density-dependent relativistic mean-field calculations,
which reproduce the empirical properties of normal nuclei with good accuracy.
We find that the evaluated NMFP due to charged current reactions significantly 
depend on the structure of the pasta phase. This is mainly due to fact that the
proton abundances derived from the $\beta$-equilibrium in inhomogeneous matter
are larger than the corresponding values determined for symmetric matter. The
effects of inhomogeneous baryonic density distributions is less pronounced for
the neutral current contribution.

Recent studies show that the weakly bound neutrons may play an important role in
formation of collective modes in the crust of neutron stars \cite{Khan04}. 
In our calculations the role of such collective features of neutrons has not yet
been considered and an accurate calculation
of nuclear response functions should be done in the future.

\section{Acknowledgments}
One of us, P. Grygorov, would like to thank Dr. V. Rodin for useful discussions
on weak interaction with nuclei. This work has been supported by the European
Graduate School ``Hadrons in Vacuum, in Nuclei and Stars'' (Basel, Graz,
T\"{u}bingen), which obtains financial support by the DFG.


\section*{References}
\begin{thebibliography}{99}

\bibitem{tubbs75} Tubbs D L and Schramm D N  1975 {\it Astrophys. J.} 
\textbf{201} 467
\bibitem{sawyer75} Sawyer R F 1975 {\it Phys. Rev.} D \textbf{11} 2740; 1989 {\it Phys. Rev.} C \textbf{40} 865

\bibitem{iwamoto82} Iwamoto N and Pethick C J 1982 {\it Phys. Rev.} D \textbf{25} 313 

%\bibitem{goodwin82} B. T. Goodwin and C. J. Pethick, Astrophys. J. \textbf{253}, 816 (1982)

\bibitem{reddy98D} Reddy S, Prakash M, and Lattimer J M 1998 {\it Phys. Rev.} D \textbf{58} 013009

\bibitem{reddy99C} Reddy S, Prakash M, Lattimer J M and Pons J A 1999 {\it Phys. Rev.} C \textbf{59} 2888

\bibitem{navarro99} Navarro J, Hernand\'{e}z E S and Vautherin D 1999 {\it Phys. Rev.} C \textbf{60} 045801 

%\bibitem{margueronPhD} J.Margueron, PhD thesis (2001) Paris, unpublished.

\bibitem{margueron03} Margueron J, Vida\~{n}a I and Bombaci I 2003 {\it Phys. Rev.} C \textbf{68} 055806 

\bibitem{ravenhall83} Ravenhall D G, Pethick C J and Wilson J R 1983 {\it Phys. Rev. Lett.} \textbf{50} 2066

\bibitem{oyamatsu93} Oyamatsu K 1993 {\it Nucl. Phys.} \textbf{A561} 431


%\bibitem{shen98} H. Shen, H. Toki, K. Oyamatsu, K. Sumiyoshi,
%{\it Nucl. Phys.} \textbf{A637}, 435 (1998).

%\bibitem{yakovlev04} D. G. Yakovlev and C. J. Pethik, Ann. Rev. Astron. Astrophys. \textbf{42}, 169 (2004).

%\bibitem{page06} D. Page, U. Geppert and F. Weber, {\it Nucl. Phys.} \textbf{A777}, 497 (2006).

%\bibitem{haxton88} W. C. Haxton, Phys. Rev. Lett. \textbf{60}, 1999 (1988).

\bibitem{Horowitz} Horowitz C J, P\'erez-Garc\'ia M A and Piekarewicz J 2004 {\it Phys. Rev.} C \textbf{69} 045804

%\bibitem{horowitz05} C. J. Horowitz, M. A. P\'erez-Garc\'ia, D. K.
%Berry, and J Piekarewicz, Phys. Rev. C \textbf{72}, 035801 (2005).

\bibitem{caballero06} Caballero O L, Berry D K and Horowitz C J 2006 {\it Phys. Rev.} C \textbf{74} 065801 

\bibitem{watanabe03} Watanabe G, Sato K, Yasuoka K and 
Ebisuzaki T 2003 {\it Phys. Rev.} C \textbf{68} 35806 

\bibitem{sonoda07} Sonoda H, Watanabe G, Sato K, Takiwaki T, Yasuoka K and Ebisuzaki T 2007
{\it Phys. Rev.} C \textbf{75} 042801  

\bibitem{Reddy} Reddy S, Bertsch G and Prakash M 2000 {\it Phys. Lett.} \textbf{B475} 1 

\bibitem{Burrows} Burrows A, Reddy S and Thompson T A 2006 {\it Nucl. Phys.} {\bf A 777} 356

\bibitem{goegel76} G\"ogelein P and M\"uther H 2007 {\it Phys. Rev.} C
\textbf{76} 024312 

\bibitem{GoegeleinPhD} G\"ogelein P 2007 {\it PhD Thesis} University of T\"{u}bingen, Germany (unpublished)

\bibitem{goegel77} G\"{o}gelein P, Van Dalen E N E, Fuchs C and M\"{u}ther H 2007 {\it Phys. Rev.} C \textbf{77} 025802  

%\bibitem{reddy99} S. Reddy, G. Bertsch, and M. Prakash, Phys. Lett. B \textbf{475}, 1 (2000).

\bibitem{Khan04} Khan E, Sandulesku N and Van Giai N 2005 {\it Phys. Rev.} C \textbf{71} 042801


%\bibitem{oyamatsu07} K. Oyamatsu and K. Iida, {\it Phys. Rev.} C
%\textbf{75}, 015801 (2007).

\bibitem{sk1} Skyrme T H R 1959 {\it Nucl. Phys.} {\bf 9} 615

\bibitem{sk2} Vautherin D and Brink D M 1972 {\it Phys. Rev.} C {\bf 5} 626

\bibitem{bv81} Bonche P and Vautherin D 1981 Nucl Phys. A{\bf 372} 496 

\bibitem{NMB:Ring80}        Ring P and Schuck P 1980
                {\it The Nuclear Many Body Problem}
                (Springer, New York) p~716

\bibitem{Chabanat98} Chabanat E, Bonche P, Haensel P, Meyer J,
Schaeffer R 1998 {\it Nucl. Phys.} {\bf A635} 231

\bibitem{TD:Bonche85}  Bonche P, Flocard H, Heenen P -H, Krieger S J,
Weiss M S 1985 {\it Nucl. Phys.} {\bf A443} 39 

\bibitem{Davies80}   Davies K T R, Flocard H, Krieger S and Weiss M S 1980
                {\it Nucl. Phys.} {\bf A342} 111

\bibitem{Montani:2004} Montani F, May C and M\"{u}ther H 2004 {\it Phys.
Rev.} C {\bf 69} 065801

%\bibitem{Kuckei03}  Kuckei J, Montani F, M\"{u}ther H and Sedrakian A 2003
%                {\it Nucl. Phys.} {\bf A723} 32

%\bibitem{DP:Bertsch91}  Bertsch G F and Esbensen H 1991 {\it Ann. Phys.} {\bf 209} 327

%\bibitem{PP:Garrido99}  Garrido E, Sarriguren P, Moya de Guerra E,
%and Schuck P 1999  {\it Phys. Rev.} {\bf C60} 064312

%\bibitem{PNM:Fayans00}  Fayans S A, Tolokonnikov S V, Trykov E L, and
%Zawischa D 2000 {\it Nucl. Phys.} {\bf A676} 49 

\bibitem{semilepton} Walecka J D 1975 {\it Muon Physics} vol. II ed
Highes V H and Wu C S (Academic, New York);
O'Connell J C, Donnelly T W, and Walecka J D 1972 {\it Phys. Rev.} C
\textbf{6} 719 ; Donnelly T W and Walecka J D 1976 {\it Nucl.
Phys.} \textbf{A274} 368 ; Donnelly T W and Haxton W C 1979
{\it Atomic Data Nucl. Data Tables} \textbf{23} 103

\bibitem{Eisenberg} Eisenberg J M and Greiner W 1970
                {\it Excitation Mechanisms of the Nucleus}
                (North-Holland Publishing Company, Amsterdam-London) p~487

\bibitem{Schiller:2001} Schiller E and M\"{u}ther H 2001 {\it Eur. Phys. J.} {\bf A11} 15

\bibitem{Hofmann:2001} Hofmann F, Keil C M and Lenske H 2001 {\it Phys. Rev.} C \textbf{64} 034314

\bibitem{vanDalen:2007} Van Dalen E N E, Fuchs C and  Faessler A 2007 {\it Eur. Phys. J.} A \textbf{31} 29

\bibitem{Klaehn:2006}Kl\"ahn T {\it et al} 2006 {\it Phys. Rev.} C \textbf{74} 035802

\bibitem{Lattimer91} Boguta J 1981 {\it Phys. Lett.} B \textbf{ 106} 255; Lattimer J M, Pethik C J, Prakash M and Haensel P 1991 {\it Phys. Lett.} \textbf{66} 201

\end{thebibliography}
\end{document}